\documentclass[prd,aps,preprintnumbers,amssymb]{revtex4}
\usepackage{graphicx}
\usepackage{epsf}
\usepackage{amsmath}
\usepackage{epstopdf}
\usepackage{multirow}
\usepackage{bm}
\usepackage{color}
\def\e3p{$\eta \rightarrow 3 \pi$}

\begin{document}
\title{%
\hfill{\normalsize\vbox{%
\hbox{}
 }}\\
{Electromagnetic potentials and radiation of the Proca field created by a massless charge }}

\author{Renata Jora
$^{\it \bf a}$~\footnote[1]{Email:
 rjora@theory.nipne.ro}}

\affiliation{$^{\bf \it a}$ National Institute of Physics and Nuclear Engineering PO Box MG-6, Bucharest-Magurele, Romania}

\date{\today}

\begin{abstract}

We study a theory less explored in the literature, that of a massive abelian gauge boson interacting with a massless fermion. In this framework we calculate the electromagnetic potentials created by a point like massless charge. We determine also  the radiation associated with a  change in the momentum of the charge in the low energy regime. The results show both similarities and discrepancies with respect to quantum electrodynamics.

\end{abstract}

\maketitle

\section{Introduction}

Quantum electrodynamics is one of the most tested  experimentally and theoretically theories \cite{Feynman1}-\cite{Peskin}. The model is based on the abelian gauge invariance in the presence of generations of fermions with quantized electrical charges. In the real world all fermions that we know have masses, some of them very small but nevertheless present.  The standard model contains not only a realized abelian gauge symmetry but also a spontaneously broken gauge symmetry encapsulated by the $Z$ gauge boson. The presence of the $Z$ boson  permits the calculations and verification of all sort of properties of the massive abelian gauge boson again in the presence of fermions.

Some properties of electrodynamics at the classical level or in the context of special theory of relativity were known early on even at the dawn of the twentieth century. An example are the Lienard-Wiechert \cite{Lienard},\cite{Wiechert} electromagnetic potentials created by a point like charge. In connection with this one may also study the electromagnetic radiation created by an accelerating or decelerating charge, the bremsstrahlung radiation.  Later on these processed were rediscussed, reinterpreted and corrected in very good agreement with the experiments in the context of the quantum electrodynamics. In quantum field theories all these processes are also related with the electromagnetic vertex, to the elastic scattering of fermions and the list might continue.

Since in real life physics most abelian gauge fields are massless and all fermions are massive more calculations were done in this framework than for alternative situations. Of course for massless Maxwell fields one may always consider the ultrarelativistic limit with very high momenta of fermions which might offer a glimpse in the situation where the fermion are massless.

In this work we plan to revisit and contribute with some knowledge to the case less discussed in the literature, that of  massive gauge abelian field of the Proca type \cite{Proca}, \cite{Stu} interacting with massless fermions. In section II we will calculate the electromagnetic potentials (of a massive gauge boson) created by a point like charge associated to a massless fermion. This will be done simply in the context of QFT and then generalized through the application of a Lorentz transformation. In section III we will discuss the radiation emitted by a charged massless fermion. Section IV is dedicated to the conclusions.

\section{Electromagnetic potentials of a massive gauge abelian boson created by a point like massless charge}

We consider the minimal Lagrangian of a massive abelian gauge field interacting with a massless fermion:
\begin{eqnarray}
{\cal L}=-\frac{1}{4}F^{\mu\nu}F_{\mu\nu}+m_A^2A^{\mu}A_{\mu}+\bar{\Psi}i\gamma^{\mu}D_{\mu}\Psi,
\label{lagr5454}
\end{eqnarray}
where $m_A$ is the mass of the gauge boson and $D_{\mu}=\partial_{\mu}+ieA_{\mu}$ is the covariant derivative.

The equation of motion for the gauge field reads:
\begin{eqnarray}
\partial_{\mu}\partial^{\mu}A^{\nu}-\partial_{\mu}\partial^{\nu}A^{\mu}+m_A^2A_{\nu}=e\bar{\Psi}\gamma^{\nu}\Psi.
\label{res5344356}
\end{eqnarray}

In the case of a Proca type of field a generalized Lorentz gauge condition works in the form $\partial_{\mu}A^{\mu}=0$ such that one may write Eq. (\ref{res5344356}) as:
\begin{eqnarray}
\partial_{\mu}\partial^{\mu}A^{\nu}+m_A^2A_{\nu}=e\bar{\Psi}\gamma^{\nu}\Psi.
\label{res53443}
\end{eqnarray}

Consider the current created by a point like charged massless fermions for a time $t>0$:
\begin{eqnarray}
j^{\nu}(x)=e \int_0^{\infty} d \tau \frac{d y^{\nu}(\tau)}{d \tau}d^4(x-y(\tau)),
\label{def4554}
\end{eqnarray}
where $y^{\nu}(\tau)$ is the trajectory in terms of a parameter $\tau$ and $\frac{d y^{\nu}(\tau)}{d \tau}$ correspond to the four velocity of the particle. For a massless particle one may state:
\begin{eqnarray}
\frac{d y^{\nu}(\tau)}{d \tau}=\frac{p^{\nu}}{p^0},
\label{expr64553}
\end{eqnarray}
where $p^{\nu}$ is is the four momenta of the particle and we consider it constant.

We substitute Eq. (\ref{expr64553}) into Eq. (\ref{def4554}) to obtain:
\begin{eqnarray}
j^{\nu}(x)=e \int_0^{\infty} d\tau \frac{p^{\nu}}{p^0}\delta(x^{\mu}-\frac{p^{\mu}}{p^0}\tau).
\label{secdef4554}
\end{eqnarray}
Then the Fourier transform of $j^{\nu}(x)$ will be:
\begin{eqnarray}
&&j^{\nu}(k)=\int d^4 x \exp[ikx]j^{\nu}(x)=
\nonumber\\
&&e \int_0^{\infty} d\tau \frac{p^{\nu}}{p^0}\delta(x^{\mu}-\frac{p^{\mu}}{p^0}\tau)\exp[ikx]=
\nonumber\\
&&e\int_0^{\infty} d\tau\frac{p^{\nu}}{p^0}\exp[i(\frac{kp}{p^0}+i\epsilon)\tau],
\label{finalcurrents5343}
\end{eqnarray}
where a factor of $\exp[-\epsilon \tau]$ where $\epsilon$ is an infinitesimal parameter was introduced to make the integral convergent. Then the integral can be performed easily and leads to:
\begin{eqnarray}
&&j^{\nu}(k)=e\int_0^{\infty} d\tau\frac{p^{\nu}}{p^0}\exp[i(\frac{kp}{p^0}+i\epsilon)\tau]
\nonumber\\
&&-e\frac{p^{\nu}}{p^0}\frac{1}{i\frac{kp}{p^0}-\epsilon}=iep^{\nu}\frac{1}{kp+i\epsilon}.
\label{current4554}
\end{eqnarray}

The equation of motion for the massive electromagnetic field may be written as:
\begin{eqnarray}
\partial^2 A^{\nu}+m_A^2A^{\nu}=j^{\nu},
\label{eqmo5776}
\end{eqnarray}
or in the Fourier space as,
\begin{eqnarray}
(-k^2+m_A^2)A^{\nu}(k)=j^{\nu}(k),
\label{four6564}
\end{eqnarray}
which leads to:
\begin{eqnarray}
A^{\nu}(k)=-\frac{1}{k^2-m_A^2}j^{\nu}(k).
\label{res5342}
\end{eqnarray}
We go back to the space time variables to obtain:
\begin{eqnarray}
A^{\nu}(x)=-\int \frac{d^4 k}{(2\pi)^4}\exp[-ikx]\frac{1}{k^2-m_A^2}j^{\nu}(k).
\label{gf756}
\end{eqnarray}
We introduce the result in Eq. (\ref{current4554}) into Eq. (\ref{gf756}) which yields:
\begin{eqnarray}
A^{\nu}(x)=-ie\int \frac{d^4 k}{(2\pi)^4}\exp[-ikx]p^{\nu}\frac{1}{k^2-m_A^2}\frac{1}{kp+i\epsilon}.
\label{finalres64539}
\end{eqnarray}

There are three poles in the integrand in Eq. (\ref{finalres64539}) in terms of $k_0$: two in $k^2-m_A^2=0$ and one in $kp+i\epsilon=0$. We place all the poles below the $x$ axis (in terms of $k_0$).  We will consider only the latest one which is below the $x$ axis and will close the contour downward to pick up the corresponding residue at $k_0=\frac{kp}{p^0}$.  Then eq. (\ref{finalres64539}) becomes:
\begin{eqnarray}
&&A^{\nu}(x)=(-ie)(2\pi i)\int \frac{d^3k}{(2\pi)^4}\frac{p^{\nu}}{p^0}\times
\nonumber\\
&&\frac{1}{(\frac{\vec{k}\vec{p}}{p^0})^2-\vec{k}^2-m_A^2}\exp[i\vec{k}\vec{x}-i\frac{\vec{k}\vec{p}}{p^0}x_0]=
\nonumber\\
&&e\int \frac{d^3k}{(2\pi)^3}\frac{p^{\nu}}{p^0}\times
\nonumber\\
&&\frac{1}{(\frac{\vec{k}\vec{p}}{p^0})^2-\vec{k}^2-m_A^2}\exp[i\vec{k}\vec{x}-i\frac{\vec{k}\vec{p}}{p^0}x_0].
\label{kindfin675}
\end{eqnarray}

The integral in Eq. (\ref{kindfin675}) is hard to solve without simplifying assumptions. We shall thus consider that the massless particle has the four momenta $(p^0,0,0,p^z)$ such that the spatial momentum is aligned on the axis $z$ and moreover that the coordinates $x_1=x_2=0$. In this case the Eq. (\ref{kindfin675}) can be written and calculated as:
\begin{eqnarray}
&&A^{\nu}(x)=e\int \frac{d^3k}{(2\pi)^3}\frac{p^{\nu}}{p^0}\times
\frac{1}{(\frac{k_zp^z}{p^0})^2-\vec{k}^2-m_A^2}\exp[ik_zz-i\frac{k_zp^z}{p^0}x_0]=
\nonumber\\
&&e\frac{1}{4\pi^2}\int_0^{\infty} d k_r k_r^2\int_0^{\pi} d \theta\sin(\theta)\frac{p^{\nu}}{p^0}\times
\nonumber\\
&&\frac{1}{\frac{[k_rp^z\cos(\theta)]^2}{p^{02}}-k_r^2-m_A^2}\exp[ik_r(z-\frac{p^zx_0}{p^0})\cos(\theta)]=
\nonumber\\
&&e\frac{1}{4\pi^2}\int _{-1}^1 d y \int_0^{\infty} d k_r k_r^2\frac{p^{\nu}}{p^0}\times
\frac{1}{\frac{[k_rp^zy]^2}{p^{02}}-k_r^2-m_A^2}\exp[ik_r(z-\frac{p^zx_0}{p^0})y]=
\nonumber\\
&&e\frac{1}{4\pi^2}\int_0^1 d y \int_0^{\infty} d k_r k_r^2\frac{p^{\nu}}{p^0}\times
\frac{1}{\frac{[k_rp^zy]^2}{p^{02}}-k_r^2-m_A^2}\exp[ik_r(z-\frac{p^zx_0}{p^0})y]+
\nonumber\\
&&e\frac{1}{4\pi^2}\int _{-1}^0 d y \int_0^{\infty} d k_r k_r^2\frac{p^{\nu}}{p^0}\times
\frac{1}{\frac{[k_rp^zy]^2}{p^{02}}-k_r^2-m_A^2}\exp[ik_r(z-\frac{p^zx_0}{p^0})y]=
\nonumber\\
&&e\frac{1}{4\pi^2}\int _0^1 d y \int_0^{\infty} d k_r k_r^2\frac{p^{\nu}}{p^0}\times
\frac{1}{\frac{[k_rp^zy]^2}{p^{02}}-k_r^2-m_A^2}\exp[ik_r(z-\frac{p^zx_0}{p^0})y]+
\nonumber\\
&&e\frac{1}{4\pi^2}\int _1^0 d y \int_0^{-\infty} d k_r k_r^2\frac{p^{\nu}}{p^0}\times
\frac{1}{\frac{[k_rp^zy]^2}{p^{02}}-k_r^2-m_A^2}\exp[ik_r(z-\frac{p^zx_0}{p^0})y]=
\nonumber\\
&&e\frac{1}{4\pi^2}\int _0^1 d y \int_0^{\infty} d k_r k_r^2\frac{p^{\nu}}{p^0}\times
\frac{1}{\frac{[k_rp^zy]^2}{p^{02}}-k_r^2-m_A^2}\exp[ik_r(z-\frac{p^zx_0}{p^0})y]+
\nonumber\\
&&e\frac{1}{4\pi^2}\int _0^1 d y \int_{-\infty}^{0} d k_r k_r^2\frac{p^{\nu}}{p^0}\times
\frac{1}{\frac{[k_rp^zy]^2}{p^{02}}-k_r^2-m_A^2}\exp[ik_r(z-\frac{p^zx_0}{p^0})y]=
\nonumber\\
&&e\frac{1}{4\pi^2}\int _0^1 d y \int_{-\infty}^{\infty} d k_r k_r^2\frac{p^{\nu}}{p^0}\times
\nonumber\\
&&\frac{1}{\frac{[k_rp^zy]^2}{p^{02}}-k_r^2-m_A^2}\exp[ik_r(z-\frac{p^zx_0}{p^0})y].
\label{interedres6564}
\end{eqnarray}
Here  we made the substitution $\cos(\theta)=y$ and in the last six lines we made simultaneously the change $y\rightarrow -y$, $k_r \rightarrow -k_r$ in some of the integrals.

In summary we write:
\begin{eqnarray}
A^{\nu}(x)=e \int_0^1 d y\int_{-\infty}^{+\infty} dk_r\frac{k_r^2}{4\pi^2}\frac{p^{\nu}}{p^0}\exp[ik_r(z-\frac{p^z x_0}{p^0})y]\frac{1}{(y^2-1)k_r^2-m_A^2}.
\label{almostfin56454}
\end{eqnarray}
The integrand has two poles in the variable $k_r$; we close the contour upwards and pick the pole above the $x$ axis (in $k_r$) at $k_r=im_A\frac{1}{\sqrt{1-y^2}}$ to get:
\begin{eqnarray}
A^{\nu}(x)=-e\frac{1}{4\pi}\frac{p^{\nu}}{p^0}m_A\int_0^1 d y \frac{1}{(1-y^2)^{3/2}}\exp[-\frac{m_Ay}{\sqrt{1-y^2}}(z-\frac{p^zx_0}{p^0})].
\label{res53442}
\end{eqnarray}
We further make the substitution $\frac{y}{\sqrt{1-y^2}}=u$ to obtain:
\begin{eqnarray}
&&A^{\nu}(x)=-em_A\frac{1}{4\pi}\frac{p^{\nu}}{p^0}\int_0^{\infty} du \exp[-m_Au(z-\frac{p^zx_0}{p^0})]=
\nonumber\\
&&e\frac{1}{4\pi}\frac{1}{z-\frac{p^zx_0}{p^0}}\frac{p^{\nu}}{p^0}.
\label{finalresgdfd}
\end{eqnarray}
Here we made the implicit assumption that $z-\frac{p^zx_0}{p^0}>0$.

To resume one may also write for our set-up:
\begin{eqnarray}
&&A^0(x)=e\frac{1}{4\pi}\frac{1}{z-\frac{p^zx_0}{p^0}}
\nonumber\\
&&A^1(x)=0
\nonumber\\
&&A^2(x)=0
\nonumber\\
&&A^3(x)=e\frac{p^z}{p^0}\frac{1}{4\pi}\frac{1}{z-\frac{p^zx_0}{p^0}}.
\label{finalres5534}
\end{eqnarray}
Here $x$ reprsents generically the four coordinates with $x_1=x_2=0$. In order to obtain a more general result we shall consider a Lorentz transformation of the space time $\Lambda^i_jx_i$. Then in the new coordinates we have:
\begin{eqnarray}
&&A^3=\Lambda^3_{\mu}A^{\mu\prime}=aA'
\nonumber\\
&&A^0=\Lambda^0_{\mu}A^{\mu\prime}=bA'
\nonumber\\
&&z=ax'
\nonumber\\
&&x^0=bx'
\nonumber\\
&&p_{\mu}'=-ap^z+bp^0=-(a-b)p^0
\nonumber\\
&&p^z=ap'
\nonumber\\
&&p^0=bp'.
\label{res54663}
\end{eqnarray}
Here $aA'$ for example indicates the four dimensional scalar product in the Minkowski space between $\Lambda^3_{\mu}=a$ and $A^{\mu\prime}=A'$. We also used the relation $p^z=p^0$ valid in our Ansatz. Note that the quantities with lower indices transform differently.

Then Eq. (\ref{finalres5534}) may be written as:
\begin{eqnarray}
&&aA'=\frac{e}{4\pi}\frac{ap}{p^0}\frac{1}{ax'-bx'}
\nonumber\\
&&bA'=\frac{e}{4\pi}\frac{1}{ax'-bx'},
\label{res64553}
\end{eqnarray}
from which we deduce,
\begin{eqnarray}
aA'=\frac{ap}{p^0}bA'.
\label{res4343}
\end{eqnarray}
Since in our approach in natural units $bp=ap$ then $aA'=bA'$. From Eq. (\ref{res54663}) we get $(a-b)=-\frac{p_{\mu}'}{p^0}$ which further leads to:
\begin{eqnarray}
aA'=-\frac{e}{4\pi}ap'\frac{1}{p'x'}
\label{res54664}
\end{eqnarray}
Since $a$ is arbitrary one determines:
\begin{eqnarray}
A^{\nu\prime}(x')=-\frac{e}{4\pi}p^{\nu\prime}\frac{1}{p'x'}.
\label{rezdsfdg}
\end{eqnarray}
This result  has the correct Lorents structure. For the case at hand $p^{\nu\prime}=E'(1,\frac{\vec{v}'}{c})$. Therefore in terms of velocities one has:
\begin{eqnarray}
A^{\nu\prime}(x')=\frac{e}{4\pi}v^{\nu\prime}\frac{1}{\vec{v}'\vec{r}'-cx^{0\prime}}
\label{res6453}
\end{eqnarray}
Note that in this case $x^{\mu\prime}$ and $v^{\mu\prime}$ are generic and arbitrary.

This is the final result for the electromagnetic potential generated by a massless charge.

\section{The radiation field}

In this section we shall consider a more comprehensive picture. We start with a massless charged fermion with momenta $q^{\mu}$  for the time $t<0$ which at time $t=0$ suffers a sudden change in momenta to $p^{\mu}$. Therefore for $t<0$ there is a retarded potential by analogy with the Lienard-Wichert potentials for massless electromagnetic field, and for $t>0$ there is  an advanced potential. In section II we considered only the advanced potential and took into account only the pole $pk=0$ situated below the $x$ axis. Her we shall analyze the full picture and also the contribution of the other poles situated below the $x$ axis, those at $k_0=\pm \sqrt{\vec{k}^2+m_A^2}$.

The derivation will take place along the same lines as in the previous section, therefore here we shall only sketch the main points.  The current created by a charge with momenta $q^{\mu}$ at $t<0$ is:
\begin{eqnarray}
j^{\nu}|_{t<0}=e\int_{-\infty}^0 d\tau \frac{q^{\mu}}{q^0}\delta^4(x-\frac{q^{\mu}}{q^0}\tau).
\label{res534}
\end{eqnarray}
Then the Fourier transform of the above expression is done along the same line with Eq. (\ref{finalcurrents5343}) and the total current would be:
\begin{eqnarray}
j^{\nu}(k)=ie\Bigg[\frac{p^{\mu}}{kp+i\epsilon}-\frac{q^{\mu}}{kq-i\epsilon}\Bigg].
\label{finalcurrent45353}
\end{eqnarray}
The electromagnetic potential becomes:
\begin{eqnarray}
&&A^{\nu}(x)=\int \frac{d^4k}{(2\pi)^4}\frac{-ie}{k^2-m_A^2}\times
\nonumber\\
&&\Bigg[\frac{p^{\nu}}{kp+i\epsilon}-\frac{q^{\nu}}{kq-i\epsilon}\Bigg].
\label{res664553}
\end{eqnarray}
The pole below the $x$ axis at $kp=0$ was already considered. The pole above the $x$ axis for $kq=0$ will give similar results. Here we will consider only the poles below the $x$ axis at $k_0=\pm \sqrt{\vec{k}^2+m_A^2}$ which give the radiation field. The contour of the integral for $k_0$ is closed below the $x$ axis and the results of the complex integration is:
\begin{eqnarray}
&&A^{\nu}(x)=\int \frac{d^3k}{(2\pi)^3}-e\frac{1}{\sqrt{\vec{k}^2+m_A^2}}\exp[-ikx]\times
\nonumber\\
&&\Bigg[\exp[-ikx]\frac{p^{\nu}}{kp+i\epsilon}-\frac{q^{\nu}}{kq-i\epsilon}+c.c\Bigg]_{k_0=\sqrt{\vec{k}^2+m_A^2}}=
\nonumber\\
&&{\rm Re}\int \frac{d^3k}{(2\pi)^3}A^{\nu}(k)\exp[-ikx]
\label{some5464}
\end{eqnarray}

Then,
\begin{eqnarray}
&&A^{\nu}(k)=-e\frac{1}{\sqrt{\vec{k}^2+m_A^2}}\times
\nonumber\\
&&\Bigg[\frac{p^{\nu}}{kp+i\epsilon}-\frac{q^{\nu}}{kq-i\epsilon}\Bigg]
\label{res534}
\end{eqnarray}

The energy of the radiation field can be extracted readily from the kinetic term of the gauge field as:
\begin{eqnarray}
&&{\rm Energy}=\int \frac{1}{2}\frac{d^3k}{(2\pi)^3}(\vec{k}^2+m_A^2)A^{\nu}(k)A_{\nu}(k)=
\nonumber\\
&&\int \frac{e^2}{2}\frac{d^3k}{(2\pi)^3}(\vec{k}^2+m_A^2)\frac{1}{\vec{k}^2+m_A^2}\times
\nonumber\\
&&\Bigg[\frac{2qp}{(kp)(kq)}-\frac{p^2}{(kp)^2}-\frac{q^2}{(qp)^2}\Bigg]
\label{energ453}
\end{eqnarray}
Of course since $q^2=p^2=0$ in the last big bracket only the first term will contribute. Note that everywhere $k_0=\sqrt{\vec{k}^2+m_A^2}$. We denote $|\vec{k}|=k$. One may rewrite Eq. (\ref{energ453}) as:
\begin{eqnarray}
&&{\rm Energy}=\frac{e^2}{2}\int_0^{\infty} dk \frac{1}{(2\pi)^3}\int d \Omega \frac{k^2}{k^2+m_A^2}\times
\nonumber\\
&&\Bigg[2(1-\vec{v}\vec{v}')\frac{1}{(1-\frac{k}{\sqrt{k^2+m_A^2}}\hat{k}\vec{v})(1-\frac{k}{\sqrt{k^2+m_A^2}}\hat{k}\vec{v}')}\Bigg].
\label{finalexp56474}
\end{eqnarray}
Here we used $q^{\mu}=E(1,\vec{v})$ and $p^{\mu}=E'(1,\vec{v}')$.

The integral in $k$ is divergent. We shall solve the integrals for $k\ll m$. We expand the integrand in series in $k$ to determine:
\begin{eqnarray}
&&{\rm Energy}=e^2\int_0^{k_{max}} dk \frac{1}{(2\pi)^3}\int d \Omega \Bigg[\frac{k^2}{m_A^2}+\frac{k^3}{m_A^3}(\hat{k}\vec{v}+\hat{k}\vec{v}')\Bigg](1-\vec{v}\vec{v}')=
\nonumber\\
&&(1-\vec{v}\vec{v}')\frac{e^2}{2\pi^2}\frac{k_{max}^3}{3m_A^2}.
\label{resc7564}
\end{eqnarray}
The  second term on the right hand side of the first line brings no contribution when integrated about the angles.

\section{Conclusions}
 In this paper we revisit a topic less discussed and explored in the literature that of a massive abelian gauge field interacting with massless fermion. The mass of the gauge boson is assumed to be of the Proca or Stueckelberg type and thus not necessary due to the spontaneous  breakdown of  a gauge symmetry.

 In this framework we discussed the abelian potentials introduced by a point like massless charge and also the radiation field associated to a change in momentum. We obtained that the energy radiated for momenta smaller than the mass of the gauge boson is maximum for the case when the the initial and final velocities are perpendicular and zero when the the two velocities are parallel.

 Our findings are in good agreement with what is known in the literature about the topic and may be useful for subjects that might appear in completions of the standard model.

\end{document}